\documentclass[runningheads]{llncs}
\usepackage{placeins}
\usepackage{graphicx}
\usepackage[title]{appendix}
\usepackage[colorlinks=true, linkcolor=blue, citecolor=blue, urlcolor=blue]{hyperref}
\usepackage{cite}
\usepackage{amsmath,amssymb,amsfonts,mathtools}
\usepackage{algorithmic}    
\usepackage{graphicx}
\usepackage{textcomp}
\usepackage{xcolor}
\usepackage{verbatim}
\usepackage{cleveref}
\usepackage[T1]{fontenc}
\usepackage{listings} 
\usepackage{array}
\usepackage{booktabs}
\usepackage{multirow}
\usepackage{footnote}
\usepackage{threeparttable}
\usepackage[inline,shortlabels]{enumitem}
\usepackage{bbding}
\usepackage{changepage}
\usepackage{tabularx,multirow,longtable,xcolor,colortbl,pifont}
\usepackage{tikz}
\usetikzlibrary{positioning, arrows.meta, shapes.geometric, calc}
\tikzset{
  block/.style = {rectangle, draw, text width=5em, text centered, minimum height=3em},
  line/.style = {draw, -Latex},
  dashedbox/.style = {rectangle, draw, dashed, minimum height=2cm, minimum width=2cm, align=center},
  cloud/.style = {draw, ellipse, node distance=3cm, minimum height=2em},
  decision/.style = {diamond, draw, text width=4.5em, text badly centered, node distance=3cm, inner sep=0pt}
}

\makeatletter
\let\old@maketitle\@maketitle
\renewcommand{\@maketitle}{
\makebox[\textwidth][c]{ 
\begin{minipage}{1.15\textwidth}
\old@maketitle \end{minipage}
}
}
\makeatother

\newcommand{\syn}[1]{\text{{\fontfamily{lmr}\selectfont{\small\textit{#1}}}}}
\newcommand{\tok}[1]{\text{{\fontfamily{lmr}\selectfont\small\textbf{#1}}}}
\newcommand{\scmd}[1]{\text{{\fontfamily{lmr}\selectfont\small\textcolor{gray}{#1}}}}

\newenvironment{lstcenter}[1][0.8\linewidth]
  {\begin{center}\begin{minipage}{#1}}
  {\end{minipage}\end{center}}

\definecolor{hgreen}{HTML}{8CB26E}
\definecolor{mgreen}{HTML}{A9C695}
\definecolor{lgreen}{HTML}{D9E7D6}
\definecolor{hred}{HTML}{AC5A54}
\definecolor{mred}{HTML}{DCADA9}
\definecolor{lred}{HTML}{F1D0CD}
\definecolor{hblue}{HTML}{738DBB}
\definecolor{mblue}{HTML}{98ADD1}
\definecolor{lblue}{HTML}{DDE8FA}
\definecolor{hgray}{HTML}{585858}
\definecolor{mgray}{HTML}{BCBCBC}
\definecolor{lgray}{HTML}{F0F0F0}
\definecolor{hyel}{HTML}{D1B765}
\definecolor{myel}{HTML}{EEDDAA}
\definecolor{lyel}{HTML}{FDF2D0}
\definecolor{hpur}{HTML}{9174A3}
\definecolor{mpur}{HTML}{C4B3CE}
\definecolor{lpur}{HTML}{DFD5E6}
\definecolor{myhgreen}{RGB}{0,102,0}

\usepackage{etoolbox}
\makeatletter
\let\llncs@addcontentsline\addcontentsline
\patchcmd{\maketitle}{\addcontentsline}{\llncs@addcontentsline}{}{}
\patchcmd{\maketitle}{\addcontentsline}{\llncs@addcontentsline}{}{}
\patchcmd{\maketitle}{\addcontentsline}{\llncs@addcontentsline}{}{}
\makeatother
\usepackage{bookmark}
\newcommand{\K}{$\mathbb{K}$}

\usepackage{color, xcolor} 

\definecolor{kterminalred}{RGB}{188,46,47}
\definecolor{knonterminalpurple}{RGB}{70, 32, 126}
\definecolor{ksyntaxgreen}{RGB}{1,86,153}
\definecolor{ksyntaxblack}{RGB}{2,15,41}
\definecolor{kcomments}{RGB}{78,137,136}
\definecolor{codegray}{rgb}{0.5,0.5,0.5}

\usepackage{listings} 
\lstdefinelanguage{K}{
	upquote=true,
	sensitive=false,
	morekeywords={syntax, configuration, rule},
	keywordstyle=\bfseries,
	comment=[l][\color{kcomments}]{//}, 	morestring=[b]",
	stringstyle=\color{gray},
    literate={"}{\textquotedbl}1,
		}

\begin{document}

\title{K-CIRCT: A Layered, Composable, and Executable Formal Semantics for CIRCT Hardware IRs}
\titlerunning{K-CIRCT}
\author{Jianhong Zhao\inst{1} \and
Jinhui Kang\inst{1} \and
Yongwang Zhao\inst{1, 2(}\Envelope\inst{)} }
\authorrunning{Zhao et al.}
\institute{School of Cyber Science and Technology, College of Computer Science and\\ Technology, Zhejiang University, China \\
\email{zhaojianhong96@gmail.com}
\email{\{kangjinhui,zhaoyw\}@zju.edu.cn}
\and
State Key Laboratory of Blockchain and Data Security, Zhejiang University, China
}

\maketitle 

\begin{abstract}
CIRCT, an open-source EDA framework akin to LLVM for software, is a foundation for various hardware description languages. Despite its crucial role, CIRCT's lack of formal semantics challenges necessary rigorous hardware verification. Thus, this paper introduces K-CIRCT, the first formal semantics in {\K} for a substantial CIRCT subset adequate for simulating a RISC-V processor. Our semantics are structured into multiple layers: (1) MLIR static semantics, which covers fundamental MLIR concepts across domains; (2) CIRCT common semantics, featuring a generic hardware model that captures key hardware features across dialects; and (3) composable and extensible semantics for specific dialects, formalized individually using a unified approach. This approach has been applied to formalize CIRCT core dialects. We validated our semantics through full-rule coverage tests and assessed its practicality using the popular RISC-V hardware design, riscv-mini.

\end{abstract}

\section{Introduction}

\textbf{CIRCT}~\cite{lenharthCIRCTLiftingHardware2021} (Circuit IR Compilers and Tools) is an open-source compiler infrastructure for hardware description languages (HDLs), inspired by LLVM's~\cite{LLVMCompilerInfrastructure} success in software. It transforms both traditional HDLs (e.g., SystemVerilog~\cite{IEEEStd18002017} and VHDL~\cite{ashendenDesignerGuideVHDL2010}) and modern languages (e.g., Chisel~\cite{Chisel} and Magma~\cite{PhanrahanMagmaMagma}) into formats for simulation, synthesis, and placement. Unlike LLVM, CIRCT introduces multiple intermediate representations (IRs) for different purposes. These IRs are defined as \textit{dialects} in \textbf{MLIR}~\cite{lattnerMLIRScalingCompiler2021} (Multi-Level Intermediate Representation), a compiler infrastructure to extend, modify, and combine dialects across domains.

Unfortunately, CIRCT lacks formal semantics, undermining formal reasoning about hardware designs. This absence prevents proving the correctness of CIRCT-based hardware designs, compilers, synthesized circuits, and simulators. Furthermore, bug-finding tools like symbolic test generators and bounded model checkers depend on definitive formal semantics to function correctly.

\noindent
\textbf{Goal and Challenges}. To tackle this pivotal issue, this paper proposes the first formal semantics for CIRCT, using the {\K} framework~\cite{FrameworkTools2023} to lay a rigorous and reliable foundation for modern hardware design. Yet, formalizing CIRCT semantics is challenging due to the following features:

\textit{1. Composability and Extensibility.} A CIRCT/MLIR program often contains multiple dialects simultaneously. This multi-dialect feature demands the composability of various dialect semantics. In contrast, current efforts in formal semantics~\cite{chenEssenceVerilogTractable2023,khanVeriFormalExecutableFormal2017,khanExecutableFormalModel2022,bourgeatEssenceBluespecCore2020,dobisVerificationChiselHardware2023,carterSMACKSoftwareVerification2016,dockinsConstructingSemanticModels2016,bangSMTBasedTranslationValidation2022} primarily focus on individual languages rather than combinations of dialect semantics. Moreover, built on MLIR, CIRCT enables the creation and modification of dialects. This requires the extensibility of dialect semantics, which is also less considered by these existing efforts.

\textit{2. Generic Hardware Features.}
CIRCT compiles various HDLs into different outputs to support hardware design. It does this by capturing the features of HDLs through multiple dialects at various abstraction levels. Each dialect is designed to capture different hardware features, including combinational and sequential logic, operation concurrency, and complexities in SystemVerilog. Therefore, beyond just formalizing these features, there is a need for a generic hardware model that can formalize key hardware features across diverse semantics of these dialects. Existing efforts~\cite{chenEssenceVerilogTractable2023,khanVeriFormalExecutableFormal2017,khanExecutableFormalModel2022,bourgeatEssenceBluespecCore2020,dobisVerificationChiselHardware2023} focus on formalizing one specific HDL and lack a generic model that addresses the complexities of various dialects.

\noindent
\textbf{Insight and Solution}. We observe that the CIRCT project can be decomposed into multiple layers. (1) MLIR, as the top layer, provides a unified compiler infrastructure across domains, including machine learning~\cite{lattnerMLIRScalingCompiler2021,bangSMTBasedTranslationValidation2022} and quantum computing~\cite{mccaskeyMLIRDialectQuantum2021}. (2) CIRCT, as the middle layer, introduces multiple dialects tailored for hardware design. (3) Specific CIRCT dialects, as the bottom layers, capture different hardware aspects at different abstraction levels. For example, \verb|hw| deals with hardware module, \verb|comb| focuses on combinational logic, \verb|seq| addresses digital sequential logic, and \verb|sv| contains SystemVerilog (SV) features to facilitate translations into SV. Our insight is that we can systematically formalize this layered structure, from MLIR through CIRCT to its specific dialects. Each upper layer is well-designed to provide composability and extensibility for the lower layers. For composability, the upper layers provide effectful functions as interfaces for individual semantics of different lower layers. These functions enable lower layers to access a shared state when they are combined, facilitating their interaction without creating dependencies in their individual semantics. For extensibility, the upper layers cover the diversity of lower layers yet are concrete enough to simplify their formalization. To evaluate our approach, we present our semantics in {\K} for executability. 

MLIR static semantics \textbf{(\S~\ref{sec:mlir-static-semantics})}  provides effectful functions for various domain semantics (e.g., machine learning and hardware) to access preprocessed and normalized information. This semantics covers fundamental MLIR concepts across domains, including generic syntax, preprocessing of type and attribute aliases, and constructing a symbol table for operations. It eases the extension of various domain semantics, allowing them to focus on their domain-specific semantics. 
CIRCT common semantics \textbf{(\S~\ref{sec:circt-common-semantics})} enables flexible combinations of various CIRCT dialect semantics via effectful functions on a generic hardware model. This model covers key hardware features across dialects to ease their formalization. Specifically, we present a bit manipulation library for combinational logic and effectful functions for sequential logic and operation concurrency.

Using these upper layers, we provide a unified approach to formalize and combine the semantics of CIRCT dialects \textbf{(\S~\ref{sec:dialect-semantics})}. This approach further deconstructs dialect semantics into their minimal semantic units (i.e., operations) and provides a step-by-step process to formalize composable semantics of individual operations. Employing this approach, we have formalized core CIRCT dialects (i.e., \verb|hw|, \verb|comb|, \verb|seq|, \verb|sv|) that other dialects can transform into or from. 

We provide a simulator and a benchmark to evaluate our semantics \textbf{(\S~\ref{sec:evaluation})}. The simulator offers a user-friendly interface for hardware designers, using our executable semantics defined in {\K}. The benchmark includes two parts: (1) A full-rule coverage test suite, crucial because official CIRCT tests do not cover all the semantic rules. (2) A practical test using \textit{riscv-mini}, a popular RISC-V design in Chisel. We compile this design into over 1000 lines of CIRCT code to ensure that our semantics are sufficient for complex designs.

In summary, we make the following main contributions:

\begin{itemize}
\vspace{-8pt}
    \item  We introduce K-CIRCT, the first semantics for CIRCT capable of simulating complex hardware designs, e.g., a RISC-V design. It enables the composability and extensibility of CIRCT dialect semantics through a layered mechanism with effectful functions. We organize the semantics into layers: starting with domain-agnostic MLIR static semantics, moving to dialect-agnostic CIRCT common semantics, and ending in specific dialect semantics. Specifically, CIRCT common semantics contains a generic hardware model to cover key hardware features, simplifying hardware-specific dialects' formalization.
    \item We implement a simulator and use it to evaluate our semantics through a full-rule coverage test suite and a popular RISC-V design, mini-riscv.
\vspace{-8pt}
\end{itemize}

K-CIRCT, including the formal semantics, a semantics-based simulator, and test suites, is available at \href{https://github.com/Stevengre/circt-semantics-public}{this GitHub repository}. 

\section{Overview}

This section introduces our formalization target, CIRCT, using a running example, and presents an overview of our formalization approach.

\subsection{A Running Example for CIRCT}

\begin{figure}[t]
\vspace{-10pt}
    \centering
    \includegraphics[height=70pt,keepaspectratio]{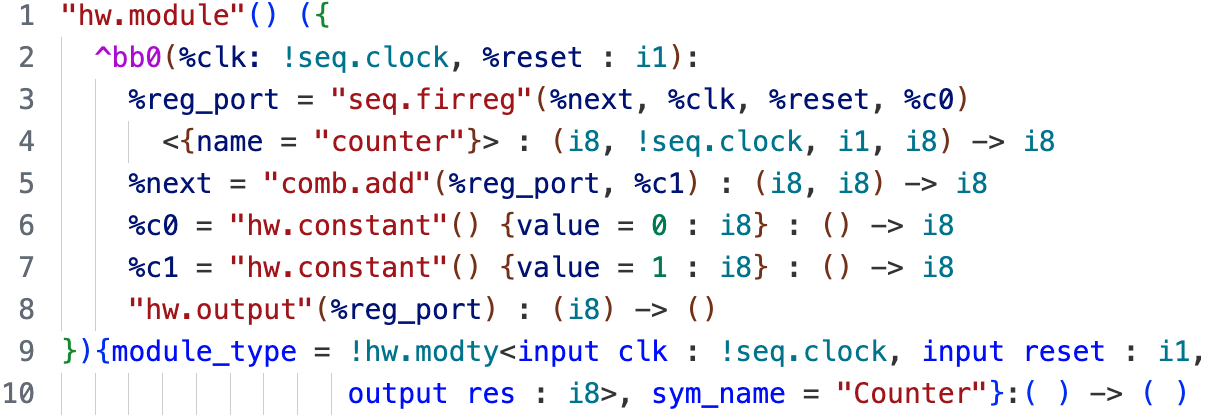}
    \hfill
    \includegraphics[height=70pt,keepaspectratio]{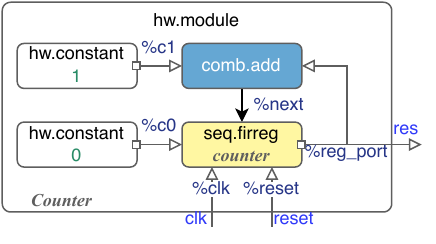}
\vspace{-10pt}
    \caption{An example of CIRCT/MLIR program.}
    \label{fig:counter-module}
\vspace{-20pt}
\end{figure}

As mentioned, MLIR is a cross-domain compiler infrastructure, and CIRCT employs this infrastructure to provide multiple dialects and combine them to describe the hardware. The dynamic semantics of operations within MLIR and CIRCT remain undefined until they are explicitly specified within a dialect. Fig.~\ref{fig:counter-module} describes the \verb|Counter| hardware circuit using a combination of dialects \verb|hw|, \verb|comb|, and \verb|seq| to show this layered semantics. Firstly, the program (Fig.~\ref{fig:counter-module} left side) conforms to the generic MLIR syntax (discussed in \S~\ref{sec:mlir-static-semantics}), with MLIR itself unaware of the meanings of specific operations. Similarly, as a hardware compiler framework, CIRCT recognizes the hardware structure and the internal component connections shown on the right side of Fig.~\ref{fig:counter-module}, yet it does not understand these components' semantic details. Clarity on the dynamic semantics of an operation emerges only when delving into specific dialects, e.g., \verb|hw.module| in \verb|hw| dialect, \verb|seq.firreg| in \verb|seq| dialect, and \verb|comb.add| in \verb|comb| dialect.

Delving deeper into this hardware description, the circuit is designed to count the clock, starting at 0 from reset. Line 1 introduces the circuit module using the \verb|hw.module| operation, with \verb|!hw.modty| specifying its interface, and \verb|sym_name| identifying its symbol in lines 9-10. A single-block region starts at line 2, connecting the input port labeled \verb|clk| to \verb|%clk|, and \verb|reset| to \verb|%reset|. Within this block, \verb|seq.firreg| presents a synchronous register, equipped with an output port for accessing stored values and four input ports for control. This synchronous setup ensures that the register's content updates only at specific edges of the clock signal. On the rising edge of the \verb|%clk| port, the register updates to a new value provided by the \verb|%next| port, and resets by the \verb|%c0| port when the \verb|%reset| port is triggered. Operation \verb|comb.add| in line 5 constructs a physical net to add the input ports \verb|%reg_port| and \verb|%c1|. Notably, the input port of \verb|comb.add| is connected to the output port of the counter register via \verb|%reg_port|. Lines 6-7 present 8-bit constant voltage nets linked to \verb|%c0| and \verb|%c1|, respectively. Finally, line 8 connects the register's output port \verb|%reg_port| to the module's output interface \verb|res|, completing the circuit's capability to start counting from 0 upon reset.

\subsection{Approach Overview}
\label{sec:overview}

\begin{figure}[ht]
\vspace{-12pt}
\centering
\includegraphics[width=0.8\textwidth]{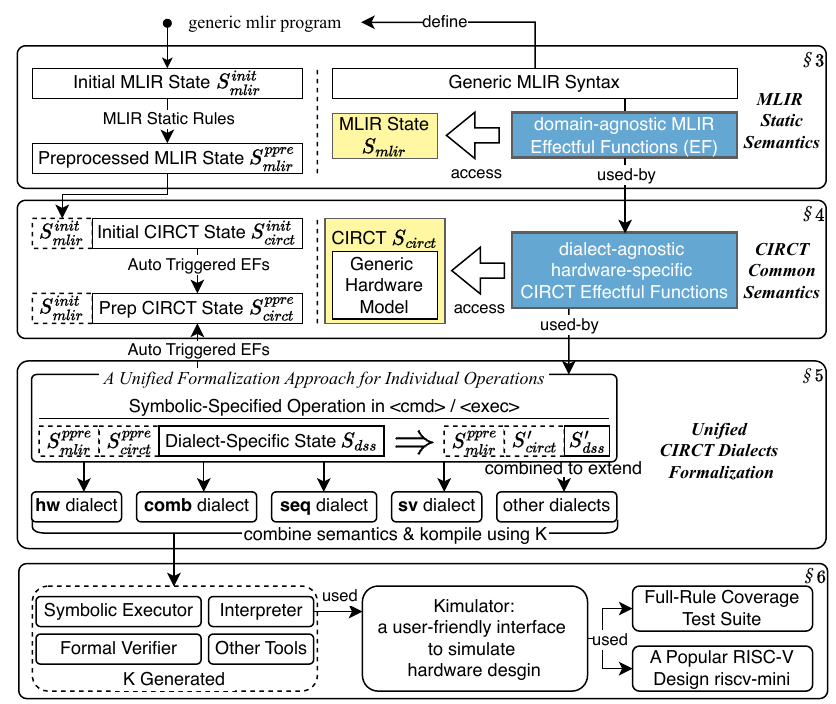}
\vspace{-14pt}
\caption{The structure of K-CIRCT.}
\label{fig:kcirct-structure}
\vspace{-20pt}
\end{figure}

To address the complexity of MLIR and CIRCT, we introduce two key strategies: (1) a layered mechanism with effectful functions for extensible and composable semantics and (2) a generic model to capture key hardware features. Fig.~\ref{fig:kcirct-structure} shows a comprehensive overview of our approach, including three basic layers of semantics. This structure progressively refines from domain-agnostic MLIR semantics to dialect-agnostic CIRCT semantics, before delving into individual semantics of various dialects. Utilizing our formalized semantics, we implement a simulator and a benchmark to evaluate the semantics.

MLIR static semantics \textbf{(\S~\ref{sec:mlir-static-semantics})} presents generic MLIR syntax to define a program, alongside an MLIR state $S_{mlir}$ to store the program and preprocessed data. The MLIR static rules just preprocess the program and store the normalized information in $S_{mlir}$ to cover the diversity of different domains. Effectful functions are interfaces for various domain-specific layers to access data in $S_{mlir}$.

CIRCT common semantics \textbf{(\S~\ref{sec:circt-common-semantics})} primarily addresses hardware features, omitting cross-domain generalizations. To house the diversity of various CIRCT dialects, the CIRCT state $S_{circt}$ contains a generic model to capture key hardware features. This model is accessible via dialect-agnostic CIRCT effectful functions, simplifying the formalization of hardware-specific semantics for dialect-specific layers. Notably, some of these effectful functions are auto-triggered, thereby eliminating the need for explicit calls by dialect semantics.

Unified CIRCT dialects formalization \textbf{(\S~\ref{sec:dialect-semantics})} presents an approach to formalize the semantics of individual operations and to combine these into specific dialect semantics. From our observation, operations like \verb|hw.module| and \verb|comb.icmp| are the minimal semantic units in CIRCT/MLIR. Each dialect comprises a set of operations that share a similar purpose and abstraction level. Other semantic elements within the dialect, including types (e.g., \verb|!hw.modty|) and attributes (e.g., \verb|0|), are designed to support the operations. Therefore, we provide a step-by-step approach to formalizing the composable semantics of individual operations for various CIRCT dialects. We then combine these operations' semantics to construct the semantics of a specific dialect. Using this method, we have formalized the semantics of core CIRCT dialects: \verb|hw|, \verb|comb|, \verb|seq|, and \verb|sv|.

By combining the semantics of these core dialects, we can use {\K}'s \textit{kompile} tool to generate several tools, including a symbolic executor, a formal verifier, an interpreter, etc. To evaluate our semantics \textbf{(\S~\ref{sec:evaluation})}, we employ the generated interpreter to implement a simulator with a user-friendly interface, called Kimulator. This simulator is then utilized for full-rule coverage testing and to check the ability of our semantics by simulating a complex hardware design \verb|riscv-mini|.

\section{Formalization of MLIR Static Semantics}
\label{sec:mlir-static-semantics}

This section focuses on formalizing domain-agnostic concepts within MLIR, enabling its cross-domain extensibility. Thus, we present: (1) a generic MLIR syntax for program description, (2) a state with effectful functions for program and information management, and (3) semantic rules for universal preprocessing.

\begin{figure}[ht]
\vspace{-26pt}
\scalebox{0.9}{
\begin{minipage}{\linewidth}
\[\begin{aligned}
&\text{TopLevel}&
\syn{toplevel} \Coloneqq &\ \scmd{(} \syn{operation} \scmd{$\mid$} \syn{ta-def} \scmd{$\mid$} \syn{aa-def} \scmd{)*}
\\
&\text{Operation}&
\syn{operation} \Coloneqq &\ \scmd{(} \syn{op-results}\ \tok{=} \scmd{)?}\ \syn{str}\ \tok{(} \scmd{(}\syn{value-uses}\scmd{)?} \tok{)}\ \scmd{(}\tok{[} \syn{successors} \tok{]}\scmd{)?}\
\scmd{(}\tok{<} \syn{dict-prop} \tok{>}\scmd{)?}
\\ 
& &
&\ \scmd{(}\tok{(} \syn{regions} \tok{)}\scmd{)?}\ \scmd{(} \tok{\{} \scmd{(}\syn{dict-attr}\scmd{)?} \tok{\}}\scmd{)?}\ \tok{:}\ \syn{fun-type}\ \scmd{(}  \syn{loc-attr} \scmd{)?}
\\
&\text{Region}& \syn{regions} \Coloneqq &\ \syn{region}\ \scmd{(} \tok{,}\ \syn{region} \scmd{)*} \hspace{51pt} \syn{region} \Coloneqq \ \tok{\{} \syn{operation}^\scmd{+}\ \syn{block}\scmd{*} \tok{\}}
\\
&\text{Block}& \syn{block} \Coloneqq &\ \syn{block-label}\ \syn{operation}^\scmd{+} \hspace{16pt} \syn{block-label} \Coloneqq \ \tok{\^}\syn{id} \ \scmd{(} \tok{(} \syn{block-args} \tok{)} \scmd{)?}\ \tok{:}
\\
&\text{Type}&
\syn{type} \Coloneqq &\ \syn{type-alias} \scmd{$\mid$} \syn{dialect-type} \scmd{$\mid$} \syn{builtin-type}
\hspace{8pt} 
\syn{ta-def} \Coloneqq \ \syn{type-alias} \ \tok{=} \ \syn{type} 
\\
&\text{Attribute}&
\syn{attr-value} \Coloneqq &\ \syn{attr-alias} \scmd{$\mid$} \syn{dialect-attr} \scmd{$\mid$} \syn{builtin-attr}
\hspace{8pt}
\syn{aa-def} \Coloneqq \ \syn{attr-alias}\ \tok{=}\ \syn{attr-value}
\\
\end{aligned}\]
\end{minipage}}
\vspace{-10pt}
\caption{Syntax for generic MLIR. Note that this figure omits some syntax definitions due to the limited space. Indeed, our formalization consists of the complete syntax. Appendix~\ref{app:generic-mlir-syntax} provides a more detailed version.}
\label{fig:mlir-syntax}
\vspace{-18pt}
\end{figure}

\textbf{Syntax}. Fig.~\ref{fig:mlir-syntax} shows (a fragment of) the syntax for the generic MLIR. Each MLIR source file contains a list of operations (\syn{operation}), type alias definitions ({\small\textit{ta-def}}), and attribute alias definitions (\syn{aa-def}). Operations serve as MLIR's unified computation unit. Identified by a unique string (\syn{str}), an operation can operate on values (\syn{value-uses}) and potentially produce a result assigned to variables (\syn{op-results}). Beyond simple instructions, operations can contain operations within \syn{regions}' blocks (\syn{block}) for complex constructs like \verb|hw.module|. Consequently, developers can integrate various representations into MLIR dialects, comprising a set of types (\syn{dialect-type}), attributes (\syn{dialect-attr}), and operations.

\begin{figure}[ht]
\vspace{-28pt}
\centering
\begin{small}
\[
\left\langle
\begin{aligned}
\langle \syn{toplevel} \rangle_{prog} \hspace{3pt}
\langle \syn{control}\rangle_{phase} \hspace{3pt}
\langle \syn{type-alias} \mapsto \syn{type} \rangle_{types}
\\
\langle \syn{attr-alias} \mapsto \syn{attr-value} \rangle_{attrs} \hspace{3pt}
\langle \syn{str} \mapsto \syn{operation} \rangle_{table} 
\end{aligned}
\right\rangle_{mlir}
\]
\end{small}
\vspace{-20pt}
\caption{The structure of MLIR state.}
\label{fig:mlir-state}
\vspace{-18pt}
\end{figure}

\textbf{State}. Fig.~\ref{fig:mlir-state} depicts the MLIR state structure, where each cell is labeled with its type inside angle brackets and its name at the bottom right. The MLIR state <mlir> establishes a generic environment, including <prog> for preprocessing programs, <phase> for execution control and pattern matching clarity, <types> and <attrs> for type and attribute aliases, and <table> for operations indexed by symbol names. Initially, the <prog> contains the MLIR source program, <phase> is set to \verb|preprocess|, and the other components within <mlir> are empty, denoted as $S_{init} :=$ \syn{<pgm>}$_{prog}$ \syn{<pre>}$_{phase}$ \syn{<>}$_{types}$ \syn{<>}$_{attrs}$ \syn{<>}$_{table}$.

\textbf{Effectful Functions}, designed to allow interaction within MLIR, enable easy access to types, attributes, and operations via aliases or symbols. These are accessible across all semantic layers through specific function calls: \syn{Rta(type-alias)} retrieves types, \syn{Raa(attr-alias)} fetches attributes, and \syn{Rop(str)} accesses operations by symbol, ensuring a unified interface for cross-layer operations.

\textbf{Semantic Rules}. Starting from the initial state $S_{init}$, the rules here preprocess programs within <prog> and store normalized details into other cells of <mlir>, achieving a preprocessed state $S_{ppre}$ that is ready for effectful functions. These rules are organized into three phases: (1) Assign Aliases (\syn{Rta} \& \syn{Raa}): This phase compiles type alias (\syn{type-alias-def}) and attribute alias (\syn{attr-alias-def}) definitions into the <types> and <attrs> cells, respectively. (2) Normalize Operations: All operations are converted to their canonical forms, eliminating concrete layers' need to manage different operation forms. (3) Construct Symbol Table (\syn{Rop}): This step decodes nested operation structures and maps operation symbols to their canonical forms in <table>. After preprocessing all operations, the phase transitions to \verb|simulation|, resulting in a preprocessed state $S_{ppre} :=$  \syn{<>}$_{prog}$\syn{<sim>}$_{phase}$\syn{<ta $\rightarrow$t>}$_{types}$\syn{<aa$\rightarrow$a>}$_{attrs}$\syn{<sid$\rightarrow$op>}$_{table}$. For example, after preprocessing for Fig.~\ref{fig:counter-module}. aliases should be empty <>$_{types}$ and <>$_{attrs}$ , and the symbol table should be <"Counter"$\mapsto$normalized line1-10>$_{table}$.

\section{Formalization of CIRCT Common Semantics}
\label{sec:circt-common-semantics}
In this section, we aim to formalize the dialect-agnostic aspects of hardware features by distilling the essential characteristics inherent to HDLs into CIRCT common semantics. This semantics is located one layer above specific dialects, enhancing cross-dialect extensibility and composability in the hardware domain.

Unlike software languages focusing on computation, CIRCT is designed to describe hardware structure, rendering the order of operations irrelevant and enabling complete concurrency. It incorporates combinational logic for managing logical gates and sequential logic to address dependencies based on historical states. To tackle these features, we formalize bit manipulation with constants X and Z for combinational logic, alongside a generic hardware model that captures circuit architecture and addresses additional hardware-domain requirements.

\begin{figure}[ht]
\vspace{-26pt}
\centering
\begin{small}
\[
\left\langle
\begin{aligned}
\langle \syn{...} \rangle_{mlir} \\
\langle \syn{cmd} \rangle_{cmd}
\end{aligned}
\left\langle
\begin{aligned}
\langle \syn{id*} \rangle_{cid} \mapsto
\langle \syn{K}\rangle^*_{exec} \hspace{3pt}
\langle \syn{id*} \rangle_{pa} \hspace{3pt}
\langle \syn{\%id} \mapsto \syn{tv} \rangle_{last} \hspace{3pt}
\langle \syn{\%id} \mapsto \syn{tv} \rangle_{curr} 
\\
\langle \syn{str} \rangle_{mod} \hspace{3pt}
\langle \syn{id} \mapsto \syn{\%id} \rangle_{out} \hspace{3pt}
\langle \syn{str} \mapsto \syn{\%id} \rangle_{reg} \hspace{3pt}
\langle \syn{str} \mapsto \syn{\%id} \rangle_{wire} 
\end{aligned}
\right\rangle_{ckt}
\right\rangle_{circt}
\]
\end{small}
\vspace{-20pt}
\caption{The structure of CIRCT common state.}
\label{fig:circt-state}
\vspace{-20pt}
\end{figure}

\textbf{Generic Hardware Model}. The common state extends the MLIR state with pending CIRCT commands <cmd> and circuit layout <ckt>, as shown in Fig.~\ref{fig:circt-state}. This layout <ckt> presents a generic hardware model that abstracts the fundamental structure of hardware without introducing specific dialects. It specifies key components like the module identifier <mod>, output ports <out>, registers <reg>, and wires <wire>. It also identifies the circuit's logical placement within a larger circuit by specifying its parent circuit module <pa>. To model ports' state for sequential logic, we introduce the current state <curr> and historical state <last>. Additionally, The circuit's internal components and their connection are defined in <exec>.

\begin{table}[ht]
\vspace{-12pt}
\centering
\caption{Effectful functions for CIRCT common state.}
\vspace{-7pt}
\begin{tabularx}{\textwidth}{llXl}
\toprule
function & Effect & Description & Trigger \\
\midrule[0.5pt]
\syn{Ec(cmd)} & <circt> & Execute a CIRCT command & \syn{cmd} in <cmd>\\
\syn{Sti(cid, vals)} & <ckt> & Stimulate a circuit with $vals$ & call in <cmd>\\
\syn{Fin} & <ckt> & Finish the simulation circuit & {<exec> auto-trigger}\\
\syn{Wags(blk,vals)} & <ckt> & Initialize the block arguments & call in <exec> \\
\syn{Rcv(\%id*)} & <curr> & Read the variables of current clock & function call \\
\syn{Wcv(\%id*, vals)} & <curr> & Write the variables of current clock & function call \\
\syn{Rlv(\%id*)} & <last> & Read the variables of lask clock & function call \\
\syn{Rreg(str)} & <reg> & Read the register by the symbol & function call \\
\syn{Wreg(str, \%id)} & <reg> & Write the register by the symbol & function call \\
\syn{Rwire(str)} & <wire> & Read the wire by the symbol & function call \\
\syn{Wire(str, val)} & <wire> & Write the wire by the symbol & function call \\
\syn{Wout(id, \%id)} & <out> & Mark the output variable & function call \\
\syn{Sop(operation)} & <prog> & Refine $operation$ in hardware domain   & <prog> auto-trigger \\
\syn{Mop(operation)} & <circt> & Move unsymboled $op$ to <cmd>  & <prog> auto-trigger \\
\syn{Sr(regions)} & <exec> & Standardize $({region})$ to $operations$   & <exec> auto-trigger \\
\syn{Pop(operations)} & <exec> & Parallelize $operations$   & <exec> auto-trigger \\
\syn{Wres(op)} & <ckt> & Assign the operation results & {<exec> auto-trigger} \\
\bottomrule
\end{tabularx}

\label{tab:circt-effect-functions}
\vspace{-20pt}
\end{table}

\textbf{Common Hardware Effectful Functions}. To enable low-coupling access to this state across hardware-domain dialect semantics, we introduce 17 effectful functions as detailed in Table~\ref{tab:circt-effect-functions}. These functions can be activated using three methods: (1) direct function calls for unrestricted usage within semantic rules; (2) constrained calls within specific cells; and (3) automatic triggers through pattern matching when the operation semantics are formalized using the standard methods (\S~\ref{sec:dialect-semantics}). With these effectful functions, we decouple dialect semantics from MLIR semantics, and address sequential logic and operation concurrency.

\textit{1. Decoupling}. For decoupling, the function \textit{Ec(cmd)} refines MLIR functions for hardware-specific applications by offering a variety of commands (\textit{cmd}). For example, the command \textit{CRop} utilizes MLIR's \textit{Rop} to retrieve operations, issuing an error for unlisted operations in <table>. The command \textit{Debug} then addresses such errors, adjusting <phase> to halt simulation.

\textit{2. Sequential Logic}. For sequential logic, we offer functions ranging from \textit{Sti} to \textit{Wout} in Table~\ref{tab:circt-effect-functions}. Specifically,  \syn{Sti(cid, vals)} is designed to stimulate a dialect-agnostic circuit with specific values within a clock cycle, extracting operations from <table> based on $cid$ and putting them into <exec>. For example, we can use \syn{Sti("Counter",0,1)} to stimulate the counter and put normalized \verb|hw.module| (lines 1-10 Fig.~\ref{fig:counter-module}) in <exec>. Following this, operations may use \syn{Wargs(blk, vals)} to assign the signals \syn{vals} to the block arguments (line 2 Fig.\ref{fig:counter-module}). With block initialization complete, operations in the block trigger their dynamic semantic rules in <exec> (lines 3-8 Fig.~\ref{fig:counter-module}), using functions like \syn{Rcv(\%id*)} for current circuit state access or \syn{Rlv(\%id*)} for previous state retrieval. Once no operation contributes to the hardware module's structure, \syn{Fin} is auto-triggered to transition register variables from <curr> to <last>.

\textit{3. Operation Concurrency and Extensibility}. For operation concurrency and extensibility, we introduce functions positioned below $Sop$ in Table~\ref{tab:circt-effect-functions} designed to manipulate operations. Initially, operations reside in <prog>, awaiting static semantic rules for canonicalization. \textit{Sop(operation)} is an auto-triggered function that refines the cross-domain normalized operation (by MLIR static rules) into the hardware domain, simplifying the format. For example, this function canonicalizes the \verb|comb.add| operation (line 5 Fig.~\ref{fig:counter-module}) into the following format:

\begin{minipage}{\linewidth}
\vspace{-3pt}
\centering
\includegraphics[width=0.8\linewidth]{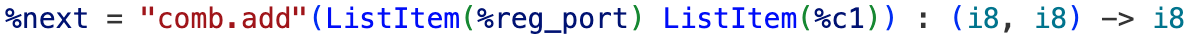}
\vspace{-3pt}
\end{minipage}

After canonicalization, an operation with a symbol name is mapped to its definition in a <table>. For operations that lack a symbol name and are not in operation with one, a seamless transition from <prog> to <cmd> is facilitated using $Mop$, allowing dialect semantics to take over. Operations are also transitioned from <table> to <exec> using $Sti$. Once <prog> is cleared, the phase transitions from \verb|preprocess| to \verb|simulation|.

\setlength{\abovedisplayskip}{-12pt}
\setlength{\belowdisplayskip}{3pt}
\begin{gather}
\frac
{}
{\langle Op\ Ops \rangle_{exec} \Rightarrow \langle Op \rangle_{exec}\ \langle Ops \rangle_{exec}}
\quad \text{\syn{Pop}}\\
\frac
{}
{\langle OpRes = PureOp \rangle_{exec} \Rightarrow \langle PureOp \sim> Wcv(OpRes) \rangle_{exec}}
\quad \text{\syn{Wres}}_0\\
\frac
{}
{\langle Signals \sim> Wcv(OpRes) \rangle_{exec} \Rightarrow \langle \rangle_{exec} \langle OpRes \mapsto Signals\rangle_{curr}}
\quad \text{\syn{Wres}}_1
\end{gather}

In the <exec> cells, operations are further simplified and parallelized to streamline the accurate formalization of hardware-specific semantics. \textit{Sr(regions)} converts a block region into a sequence of operations (i.e., hardware components), which are then split into individual parallelizable simulation cells <exec> using \syn{Pop}, obtaining individual hardware components. Subsequently, \textit{Wres} is triggered automatically to connect these components. Specifically, \textit{Wres}$_0$ sets up the components' input ports while \textit{Wres}$_1$ sets the output ports. For instance, \verb|comb.add| exposes the input ports \verb|%reg_port| and \verb|%c1| by \textit{Wres}$_0$ and the output ports \verb|%next| by \textit{Wres}$_1$ (line 5 Fig.~\ref{fig:counter-module}).

As a result, CIRCT common semantics simplifies the formalization and combination of hardware-specific dialect semantics, allowing their concentration on unique dialect features, disregarding aspects such as combinational logic, sequential logic, and operation concurrency.

\section{Unified CIRCT Dialects Formalization}
\label{sec:dialect-semantics}

Employing CIRCT common semantics, this section introduces our approach to define and combine the semantics of various CIRCT dialects. We achieve this through a unified formalization approach to defining the semantics of dialects' minimal semantic units (individual operations) and combining them in a specific dialect. Utilizing this approach, we formalize the CIRCT core dialects (\verb|hw|, \verb|comb|, \verb|seq|, \verb|sv|) and combine them to support a CIRCT/MLIR program (like Fig.~\ref{fig:counter-module}).

\setlength{\abovedisplayskip}{-5pt}
\setlength{\belowdisplayskip}{3pt}
\begin{gather}
\frac
{\langle pre \rangle_{phase}}
{\langle Op\ \dots \rangle_{cmd}\ S_{dss}\Rightarrow  \langle \rangle_{cmd}\ S_{dss}^{\prime}}
\quad Op_{pre}\\
\frac
{\langle sim \rangle_{phase}}
{\langle Op\ \dots \rangle_{exec}\ S_{dss}\Rightarrow  \langle Signals\ \dots  \rangle_{exec}\ S_{dss}^{\prime}}
\quad Op_{sim}
\end{gather}

\noindent
\textbf{A Unified Formalization Approach for Individual Operations}. The key to formalizing individual operations is to specify their symbolic form in the pending state (<cmd> or <exec>) and employ effectful functions to access the upper layers' state for composability (Fig.~\ref{fig:kcirct-structure}). The dialect-specific state $S_{dss}$ is extended on need. $Op_{pre}$ and $Op_{sim}$ are unified formalizations of various operations of CIRCT dialects. Note that we omit the transition of states $S_{mlir}$ and $S_{circt}$ in them, because the transition is triggered by the effectful functions in the upper layers. Specifically, $Op_{pre}$ represents a unified formalization of the unsymboled operation outside a block. Because the \verb|=| symbol has different semantics in different dialects, we leave its definition to their specific semantics, resulting in empty $S_{cmd}$ ($<>_{cmd}$). Similarly, $Op_{sim}$ is a unified formalization for other operations that may produce $Signals$ to their output ports.

Delving into the unified formalization process involves up to 7 steps: (1) select the phase, (2) retrieving dialect-specific values, (3) compute values according to the \textit{operation}, (4) control the simulation flow for operations with regions, (5) normalize values, (6) update the circuit state if necessary, and (7) return results. We illustrate this process through the example of \verb|comb.add| (Fig.~\ref{fig:comb-add}). 

\begin{figure}[ht]
\vspace{-20pt}
\begin{lstcenter}[0.8\linewidth]
\begin{lstlisting}[basicstyle=\scriptsize, language=K, escapeinside={(*}{*)}]
rule <phase> simulation </phase> // 1. choose phase
<exec>
  // symbolic-specified operation (hardware component) to be formalized
  "comb.add"((*\colorbox{lblue}{As}*)) : ((*\colorbox{lblue}{Ts}*)) -> ((*\colorbox{lblue}{To}*)) 
     // 7. return values, e.g., ListItem(TV(X, Y))
  => ListItem( TV((*\colorbox{lblue}{To}*), Bit2Int( // 5. normalize values, e.g., Bit2Int(X)
                add( // 3. operate on values, e.g., add(X, Y, ...)
                      // 2. get dialect-specific values, bits(vals, types)
                      bits(Rcv((*\colorbox{lblue}{As}*)), getWidth((*\colorbox{lblue}{Ts}*)))))))
... </exec>
\end{lstlisting}
\end{lstcenter}
\vspace{-12pt}
\caption{The $\mathbb{K}$ definition for comb.add.}
\vspace{-20pt}
\label{fig:comb-add}
\end{figure}

\textit{1. Phase Selection}. The first step is to select a phase based on the operation's attributes and location. Top-level operations lacking symbol names are designated the ``preprocess'' phase, e.g., \verb|sv.def|. Conversely, operations like \verb|hw.module|, despite being top-level yet possessing a symbol name, along with operations like \verb|comb.add|, that lack a symbol name yet are located in a block, are assigned to the ``simulation'' phase (line 1).

\textit{2. Retrieving Dialect-Specific Values}. This step aims to extract the operation's operands and attributes that are unique to its dialect, with attributes defining the operational logic. For example, in line 9, operand \verb|As| are read using $Rcv$, and their values are converted to the \verb|comb| dialect's bit form using \verb|bits|.

\textit{3. Operating on Values}. Operations represent different hardware components. Specifically, \verb|comb.add| is a logical adder designed to sum signals \verb|As| (line 7).

\textit{4. Simulation Flow Control}. Operations, like \verb|hw.module| and \verb|sv.if|, that involve regions represent specific control circuits. For \verb|sv.if|, which includes a conditional and a consequent one-block region, evaluating the condition and then linking to the proper circuit block. Conversely, \verb|comb.add|, lacking a region, is not this type of circuit (line 4).

\textit{5. Value Normalization}. The diversity of types in CIRCT, ranging from built-in and dialect types to attribute types, introduces complexity. To address this, we standardize these varying types into a uniform format. For \verb|comb|, we convert the Bit value to the logical integer (line 6).

\textit{6. State Updates}. Operations can have side effects, potentially altering states following computation. For example, the \verb|hw.output| operation updates the <out> state of the current instance by triggering the \syn{Wout} function.

\textit{7. Returning Values}. Operations generally produce return values formatted as a list (line 6). If an operation returns nothing, it yields an empty list \verb|.List|. These values are then assigned to ports \syn{op-results} by the prepared \syn{Wres}.

Following these steps, we can formalize arbitrary operations by specifying their name in <exec> or <cmd>. These individual definitions of operations are composable by the indirect interactions using shared states with effectful functions. Then, keeping this we can formalize the dialect semantics by combining proper operation definitions. Without losing this composability, we are also free to combine the semantics of various dialects.

\noindent
\textbf{CIRCT Core Dialects}. Utilizing our approach, we formalized CIRCT core dialects: (1) hw for general hardware representation, (2) comb for combinational logic, (3) seq for digital sequential logic, and (4) sv for SystemVerilog-specific constructs in an AST format. Fig.~\ref{fig:dialect-state} illustrates the individual declarations of these dialects' states and their convenient combination through a top-level state <top>. Each dialect-specific state, such as <force> for handling \verb|sv.force| in the \verb|seq| dialect, records unique and independent information on need. Besides the on-need extended state, auxiliary semantic rules for dialects' types and attributes are also extended on need, because they only show their meaning via operations. Additionally, we do not provide a unified mechanism for types and attributes, because the auxiliary rules for them are pure and easy to combine.
\begin{figure}[ht]
\vspace{-28pt}
\centering
\begin{small}
\[
\left\langle
\begin{aligned}
\langle \syn{\dots} \rangle_{log} \hspace{3pt}
\langle
\langle \syn{id*} \rangle_{finst} \mapsto
\langle \syn{\%id} \mapsto tv \rangle_{fvars} 
\rangle_{force}
\\
\langle \syn{str} \mapsto \syn{str}  \rangle_{macro} \hspace{3pt}
\langle \syn{str*} \rangle_{inited} \hspace{3pt}
\langle \syn{str} \mapsto \syn{str} \rangle_{fd}
\end{aligned}
\right\rangle_{sv}
\left\langle
\begin{aligned}
\langle \syn{str} \mapsto \syn{hw.heir*} \rangle_{hier}
\\
\langle \syn{hw.heir} \mapsto \syn{id} \rangle_{h2inst}
\end{aligned}
\right\rangle_{hw}
\]
\vspace{-30pt}
\\
\[
\langle \langle \syn{\dots} \rangle_{mlir} \hspace{3pt}
\langle \syn{\dots} \rangle_{circt} \hspace{3pt}
\langle \syn{\dots} \rangle_{sv} \hspace{3pt}
\langle \syn{\dots} \rangle_{hw} \hspace{3pt}
\dots
\rangle_{top}
\]
\end{small}
\vspace{-28pt}
\caption{The structure of dialect states and their combination.}
\label{fig:dialect-state}
\vspace{-20pt}
\end{figure}

\section{Evaluation}
\label{sec:evaluation}

\subsection{Statistics of Semantics}
\label{sec:statistics}

We established the semantics for 98 CIRCT operations, with statistics provided in Table~\ref{tab:statistics}. This table categorizes various metrics such as source lines of code (SLOC), comment lines of code (CLOC), rule count (N rules) and operation count (N ops) across its rows. The columns detail the MLIR semantics, CIRCT common semantics, and the dynamic semantics for hw, comb, seq, and sv dialects, leading to a summary of the total size. For a complete list of the formalized operations, please refer to the Appendix~\ref{app:list-of-operations}.

\begin{table}[!h]
\vspace{-12pt}
\centering
\caption{Statistics for our semantics.}
\begin{tabular}{cccccccc}
\toprule
& MLIR & CIRCT & HW & COMB & SEQ & SV & Total \\ \midrule[0.5pt]
SLOC    & 555 & 547 & 582 & 351 & 68 & 1190 & 3969 \\ 
CLOC    & 91 & 31 & 21 & 35 & 13 & 58 & 346 \\ 
       N rules & 61 & 63 & 54 & 44 & 6 & 111 & 461 \\ 
       N ops & - & - & 20 & 20 & 5 & 53 & 98 \\ 
\bottomrule
\end{tabular}
    
    \label{tab:statistics}
    \vspace{-28pt}
\end{table}

\subsection{Kimulator: Formal Semantics-Based Simulation}
\label{sec:simulation}

\begin{figure}[ht]
\vspace{-12pt}
    \centering
    \includegraphics[width=0.8\linewidth]{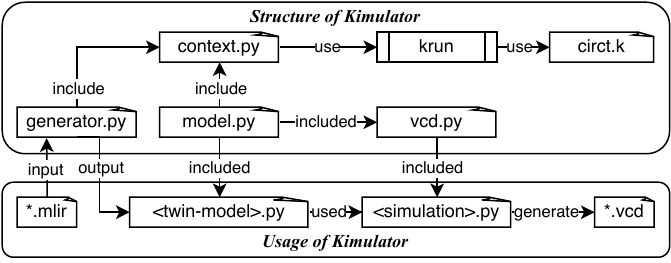}
    \vspace{-12pt}
    \caption{The architecture of Kimulator.}
    \label{fig:kimulator-structure}
\vspace{-20pt}
\end{figure}

This section introduces Kimulator, a simulation tool that employs a Python interface akin to Verilator, enhancing usability with our semantics. Fig.~\ref{fig:kimulator-structure} depicts Kimulator's architecture, illustrating both its usage and structure.

\textbf{Usage of Kimulator}.
Kimulator simplifies interaction with CIRCT semantics through \textit{generator.py}, a pivotal component for initiating simulation. It parses CIRCT source files, generating corresponding Python hardware modules to utilize CIRCT formal semantics within the $\mathbb{K}$ framework. Within these modules, \textit{model.py} provides functionalities for manipulating hardware modules and activating semantics, enabling users to develop simulation scripts in Python. These scripts can repeatedly stimulate hardware modules, allowing for simulation result analysis via the generated VCD files with \textit{vcd.py}. Compared to the famous open-source simulator Verilator, Kimulator offers a streamlined simulation experience without increasing the workload, despite employing a formal framework. For a detailed comparison, please refer to the illustration in Appendix~\ref{app:comparison-kimulator-varilator}.

\textbf{Structure of Kimulator}. Diving into Kimulator’s structure, \textit{generator.py} generates instances of the \textit{Signal} and \textit{KimulatorModel} classes, along with a \textit{KimulatorContext} instance to encapsulate the global simulation information. \textit{context.py} provides the \textit{Signal} class, which structures the hardware module, and \textit{KimulatorContext}, which manages the simulation state and interfaces with the $\mathbb{K}$ framework. \textit{model.py} introduces the \textit{KimulatorModel} class, mirroring hardware modules in Python and leveraging \textit{KimulatorContext}'s functions to execute simulations via \textit{krun}, based on the \textit{kompile}d \textit{circt.k}. The tools \textit{krun} and \textit{kompile}, provided by the $\mathbb{K}$ framework, are essential for the interpretation and tool generation, respectively. To synchronize changes between the Python twin model and the formal circuit representation, modifications and data extractions are performed on \textit{krun}'s output through Kore, an intermediate representation in $\mathbb{K}$. This approach ensures that execution can continue seamlessly into the next simulation cycle, allowing for the dynamic update of the Python model post-simulation. Lastly, \textit{vcd.py} serves to store simulation outcomes in a VCD file, streamlining result analysis and storage.

In conclusion, Kimulator offers a user-friendly simulation platform using our semantics, reducing the complexity of formal methods and the $\mathbb{K}$ framework.

\subsection{Validation of Semantics}
\label{sec:validation}

Using Kimulator, this section provides the approach for building confidence in our semantics and proposes several findings based on our validation results.

\begin{figure}
\vspace{-7pt}
    \centering
    \includegraphics[width=0.8\textwidth]{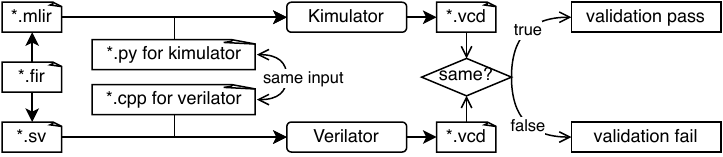}
    \vspace{-12pt}
    \caption{Co-simulation against verilator for semantics validation.}
    \label{fig:co-simulation}
    \vspace{-20pt}
\end{figure}

Fig.~\ref{fig:co-simulation} illustrates our validation strategy by co-simulation with a popular Verilog simulator, i.e., Verilator. Given the same chisel, firrtl, or other hardware design source file using CIRCT, we utilize the CIRCT project to compile them into generic MLIR or Verilog, which can be simulated by Kimulator and Verilator, respectively. Then, we give the same inputs to stimulate the same hardware (e.g., Fig.~\ref{fig:kimulator-example}). After each simulation cycle, the output is captured in VCD files, allowing for a detailed comparison. Discrepancies between the Kimulator and Verilator outputs prompt a thorough error analysis, potentially attributed to various factors, including documentation ambiguities or errors in the formal semantics implementation, Kimulator, Verilator, or the stimulation source files.

Our experiments are divided into two parts:  (1) Operation-level validation, for validating our semantics by testing all the semantic rules and formalized operations, and (2) Hardware-level validation, for assessing the practicality of our formalized operations by simulating a popular RISC-V design, \verb|riscv-mini|.

\textbf{Operation-Level Validation}. The goal here is to validate our semantics through a full-rule coverage test suite, that contains 86 CIRCT/MLIR programs and doubled simulation sources for both Kimulator and Verilator. The inputs for these programs are meticulously designed to ensure full-rule coverage and are stored in a file to provide the same inputs for co-simulations. The design of the test suite involves the following steps:

\textit{1. Initial Test Selection}: Initially, we considered using existing tests from Chisel hardware designs, but their complex logic obscured the semantic effects we sought to observe. We then evaluated the existing tests within the CIRCT project, which are primarily compilation-oriented and do not adequately cover all the semantic rules. For example, \href{https://github.com/llvm/circt/tree/172b5eabb5b2c6badde548a38f09bbb9a81d3277/test}{the test suites} of CIRCT don't cover all the events of the \verb|sv.alwaysff| like \verb|negedge|. The complexity of CIRCT's tests also hindered rapid validation and error identification.

\textit{2. Test Adaptation and Augmentation}: We extracted and modified tests from the CIRCT project, supplementing them with additional tests designed for full-rule coverage. To aid in rule-coverage determination and accelerate testing, we designed each program to focus on a minimal set of rules, typically validating the semantics of one operation per program. 
\textit{3. Compiler Compatibility Check}: All programs were checked for compilability via \verb|circt-opt| to align with the compiler’s expectations and to ensure no errors or ambiguities caused by the CIRCT documentation.

\textit{4. Handling Operation Interdependencies}: Despite efforts to isolate operation semantics, the inherent interrelations among CIRCT dialect operations (at compiler level) necessitated their combination for co-simulation. For example, operations within the \verb|comb| dialect cannot exist outside of \verb|hw.module|, and \verb|sv.force| continues to override procedural assignments until \verb|sv.release| is executed. We identified these interrelations and made minimal adjustments to the programs, such as modifying \verb|comb.add| to \verb|comb.xor| to progressively cover all the rules. 

\textit{5. Input Design}: We specifically designed test inputs for operations sensitive to values, to ensure each rule is invoked. For instance, the \verb|comb.icmp| operation utilizes the \verb|predicate| attribute to define ten comparison modes, each formalized by a separate rule. We achieved complete coverage by assigning ten distinct values to \verb|predicate|, one for each mode.

\textit{6. Simulation Source Adjustments}: We simulated multiple clock cycles to validate rules and operations requiring access to the historical state <last>.

\smallskip
\textit{Results}. Our semantics successfully passed all operation-level tests, affirming its correctness. However, several issues were identified, including:

\begin{itemize}
\vspace{-7pt}
\item We have found over 20 flaws in the MLIR documentation, such as missing definitions and inconsistent syntax. These issues have acknowledged by the LLVM community, as recorded in GitHub Issues \href{https://github.com/llvm/llvm-project/issues/62489}{62489}, \href{https://github.com/llvm/llvm-project/issues/62488}{62488}, \href{https://github.com/llvm/llvm-project/issues/62490}{62490}. 
\item Additionally, we have identified numerous flaws in the CIRCT documentation. The multitude of issues prevents us from formalizing and submitting patches simultaneously. Even worse, we must repeatedly check the legality of operations accepted by the CIRCT implementation and compile them into Verilog to infer their semantics.

\item The documentation for CIRCT version changes is lacking. However, dialects change across CIRCT versions and affect their semantics, such as attributes, module type notations, and built-in instance unfolding methods. Although our semantics is extensible, checking these changes requires substantial time.
\vspace{-18pt}
\end{itemize}

\textbf{Hardware-Level Validation}. The goal here is to validate the practical applicability of our formalized operations for complex hardware design, such as the RISC-V design. We opted for the riscv-mini project, a popular RISC-V 3-stage pipeline in Chisel. Despite its simplicity, it consists of significant features of a processor, including RV32I from the User-level ISA Version 2.0 and the Machine-level ISA from the Privileged Architecture Version 1.7, as well as basic instruction and data caches. The riscv-mini, compiled to 1455 lines of CIRCT code, serves as an ideal testbed to determine if our formalized operations adequately cover the functional requirements of real-world hardware descriptions. Through multiple input tests on the top module, we validate our semantics.

\smallskip
\textit{Results}.  (1) This validation phase polished our semantics by revealing undocumented operation attributes. (2) We observed that certain operations compiled at the hw-level were underutilized, indicating potential areas for further capturing the essence of CIRCT.

\section{Related Work}

To the best of our knowledge, our work pioneers the formal semantics for CIRCT, introducing precision and determinism to the unified hardware compiler framework. This section compares our contributions with prior efforts in formal semantics that overlap in scope (i.e., MLIR) or target similar objectives (i.e., HDLs). 
Bang et al.'s seminal work on MLIR dialects~\cite{bangSMTBasedTranslationValidation2022} encodes memref, linalg, and tensor dialects as a whole in SMT to enable translation validation for machine learning compilers. While insightful, it diverges from our scope by concentrating on a different subset of dialects without addressing the extensibility and composability that our approach emphasizes. Further, the proposal by  Yu~\cite{yuReasoningMLIRSemantics2023} acknowledges the challenges of composability in MLIR formalization, suggesting the adoption of effect handlers as a novel solution. However, his proposal, merely outlining a simple idea, underestimates the complexities inherent in real-world projects. Our methodology leverages insights from their approach but advances further by employing a layered mechanism in our semantics. This innovation allows for greater extensibility and composability, addressing the complexities overlooked in the preliminary proposal.

The exploration of formal semantics within the HDL domain has traditionally been isolated, benefiting single languages and lacking the cross-language power found in software formalization efforts like SMACK~\cite{baranowskiVerifyingRustPrograms2018,kalraZeusAnalyzingSafety2018,carterSMACKSoftwareVerification2016,rakamaricSMACKDecouplingSource2014}, and SAW~\cite{dockinsConstructingSemanticModels2016}. Specific to HDLs, contributions such as \cite{chenEssenceVerilogTractable2023,khanVeriFormalExecutableFormal2017,wolfYosysaFreeVerilog2013,meredithFormalExecutableSemantics2010} for Verilog, \cite{khanExecutableFormalModel2022} for VHDL, \cite{bourgeatEssenceBluespecCore2020} for BlueSpec, and \cite{dobisVerificationChiselHardware2023} for Chisel, showcase the depth of work within individual languages. This fragmented approach may stem from the lack of a unified intermediate representation (IR) for hardware, contrasting the cohesive LLVM IR in software. Despite HDLs' propensity for generating Verilog to tap into its established ecosystem, Verilog's primary orientation towards circuit synthesis limits its efficacy as a unifying IR. Recognizing this gap, the LLVM community's introduction of CIRCT seeks to transpose successful practices from software to hardware. Our contribution marks a critical step in this transformative journey, offering the first formal semantics for CIRCT that paves the way for future unified verification of hardware designs.

\section{Conclusion}

We introduced K-CIRCT, the first formal semantics for CIRCT, capable of simulating a popular RISC-V design. Through a layered mechanism and effectful functions, we provide a unified approach to define and combine dialect semantics. Specifically, the MLIR static semantics simplifies the formalization of MLIR-based projects across various domains. The CIRCT common semantics streamlines the formalization of hardware-domain dialects by introducing a generic hardware model. Utilizing this unified approach, we have successfully formalized the key dialects—\verb|hw|, \verb|comb|, \verb|seq|, and \verb|sv|. Our semantics have been tested through a full-rule coverage test suite and the simulation of the popular RISC-V design, \verb|riscv-mini|, supported by a user-friendly simulator that we developed.

Leveraging the {\K} framework, we can derive various formal analysis tools from our semantics, e.g., a symbolic interpreter and an equivalence checker. These tools are useful for hardware symbolic execution, formal verification, equivalence verification, and translation validation, significantly enhancing the trustworthiness of hardware designs. Future efforts will focus on collaborating with the CIRCT and {\K} communities to enhance scalability and further integrate formal methods into the development of CIRCT tools.

\bibliographystyle{splncs04.bst}
\bibliography{main}

\begin{thebibliography}{10}
\providecommand{\url}[1]{\texttt{#1}}
\providecommand{\urlprefix}{URL }
\providecommand{\doi}[1]{https://doi.org/#1}

\bibitem{Chisel}
Chisel, \url{https://www.chisel-lang.org/}

\bibitem{IEEEStd18002017}
{{IEEE Std}} 1800-2017 ({{Revision}} of {{IEEE Std}} 1800-2012) {{IEEE
  Standard}} for {{SystemVerilog}}---{{Unified Hardware Design}},
  {{Specification}}, and {{Verification Language}}

\bibitem{LLVMCompilerInfrastructure}
The {{LLVM Compiler Infrastructure Project}}, \url{https://llvm.org/}

\bibitem{PhanrahanMagmaMagma}
Phanrahan/magma: Magma circuits, \url{https://github.com/phanrahan/magma}

\bibitem{FrameworkTools2023}
K {{Framework Tools}}. Runtime Verification Inc. (Apr 2023),
  \url{https://github.com/runtimeverification/k}

\bibitem{ashendenDesignerGuideVHDL2010}
Ashenden, P.J.: The Designer's Guide to {{VHDL}}. Morgan Kaufmann (2010)

\bibitem{bangSMTBasedTranslationValidation2022}
Bang, S., Nam, S., Chun, I., Jhoo, H.Y., Lee, J.: {{SMT-Based Translation
  Validation}} for {{Machine Learning Compiler}}. In: Computer {{Aided
  Verification}}. pp. 386--407. Springer, Cham (2022),
  \url{https://linkspringer.53yu.com/chapter/10.1007/978-3-031-13188-2_19}

\bibitem{baranowskiVerifyingRustPrograms2018}
Baranowski, M., He, S., Rakamari{\'c}, Z.: Verifying {{Rust}} programs with
  {{SMACK}}. In: International {{Symposium}} on {{Automated Technology}} for
  {{Verification}} and {{Analysis}}. pp. 528--535. Springer (2018)

\bibitem{bourgeatEssenceBluespecCore2020}
Bourgeat, T., {Pit-Claudel}, C., Chlipala, A., {Arvind}: The essence of
  {{Bluespec}}: A core language for rule-based hardware design. In: Proceedings
  of the 41st {{ACM SIGPLAN Conference}} on {{Programming Language Design}} and
  {{Implementation}}. pp. 243--257. ACM, London UK (Jun 2020),
  \url{https://dl.acm.org/doi/10.1145/3385412.3385965}

\bibitem{carterSMACKSoftwareVerification2016}
Carter, M., He, S., Whitaker, J., Rakamaric, Z., Emmi, M.: {{SMACK}} software
  verification toolchain. In: 2016 {{IEEE}}/{{ACM}} 38th {{International
  Conference}} on {{Software Engineering Companion}} ({{ICSE-C}}). pp.
  589--592. IEEE (2016)

\bibitem{chenEssenceVerilogTractable2023}
Chen, Q., Zhang, N., Wang, J., Tan, T., Xu, C., Ma, X., Li, Y.: The {{Essence}}
  of {{Verilog}}: {{A Tractable}} and {{Tested Operational Semantics}} for
  {{Verilog}}. Proceedings of the ACM on Programming Languages
  \textbf{7}(OOPSLA2),  234--263 (Oct 2023),
  \url{https://dl.acm.org/doi/10.1145/3622805}

\bibitem{dobisVerificationChiselHardware2023}
Dobis, A., Laeufer, K., Damsgaard, H.J., Petersen, T., Rasmussen, K.J.H.,
  Tolotto, E., Andersen, S.T., Lin, R., Schoeberl, M.: Verification of {{Chisel
  Hardware Designs}} with {{ChiselVerify}}. Microprocessors and Microsystems
  \textbf{96},  104737 (Feb 2023),
  \url{https://linkinghub.elsevier.com/retrieve/pii/S0141933122002666}

\bibitem{dockinsConstructingSemanticModels2016}
Dockins, R., Foltzer, A., Hendrix, J., Huffman, B., McNamee, D., Tomb, A.:
  Constructing {{Semantic Models}} of {{Programs}} with the {{Software Analysis
  Workbench}}. In: Blazy, S., Chechik, M. (eds.) Verified {{Software}}.
  {{Theories}}, {{Tools}}, and {{Experiments}}, vol.~9971, pp. 56--72. Springer
  International Publishing, Cham (2016),
  \url{http://link.springer.com/10.1007/978-3-319-48869-1_5}

\bibitem{kalraZeusAnalyzingSafety2018}
Kalra, S., Goel, S., Dhawan, M., Sharma, S.: Zeus: Analyzing safety of smart
  contracts. In: Ndss. pp. 1--12 (2018)

\bibitem{khanExecutableFormalModel2022}
Khan, W., Hou, Z., Sanan, D., Nebhen, J., Liu, Y., Tiu, A.: An {{Executable
  Formal Model}} of the {{VHDL}} in {{Isabelle}}/{{HOL}} (Feb 2022),
  \url{http://arxiv.org/abs/2202.04192}

\bibitem{khanVeriFormalExecutableFormal2017}
Khan, W., Tiu, A., San{\'a}n, D.: {{VeriFormal}}: {{An Executable Formal
  Model}} of a {{Hardware Description Language}}. In: {{SG-CRC}}. pp. 19--36
  (2017)

\bibitem{lattnerMLIRScalingCompiler2021}
Lattner, C., Amini, M., Bondhugula, U., Cohen, A., Davis, A., Pienaar, J.,
  Riddle, R., Shpeisman, T., Vasilache, N., Zinenko, O.: {{MLIR}}: {{Scaling
  Compiler Infrastructure}} for {{Domain Specific Computation}}. In: 2021
  {{IEEE}}/{{ACM International Symposium}} on {{Code Generation}} and
  {{Optimization}} ({{CGO}}). pp. 2--14. IEEE, Seoul, Korea (South) (Feb 2021),
  \url{https://ieeexplore.ieee.org/document/9370308/}

\bibitem{lenharthCIRCTLiftingHardware2021}
Lenharth, A., Lattner, C.: {{CIRCT}}: {{Lifting}} hardware development out of
  the 20th century  (2021)

\bibitem{mccaskeyMLIRDialectQuantum2021}
McCaskey, A., Nguyen, T.: A {{MLIR Dialect}} for {{Quantum Assembly
  Languages}}. In: 2021 {{IEEE International Conference}} on {{Quantum
  Computing}} and {{Engineering}} ({{QCE}}). pp. 255--264. IEEE, Broomfield,
  CO, USA (Oct 2021), \url{https://ieeexplore.ieee.org/document/9605269/}

\bibitem{meredithFormalExecutableSemantics2010}
Meredith, P., Katelman, M., Meseguer, J., Rosu, G.: A formal executable
  semantics of {{Verilog}}. In: Eighth {{ACM}}/{{IEEE International
  Conference}} on {{Formal Methods}} and {{Models}} for {{Codesign}}
  ({{MEMOCODE}} 2010). pp. 179--188. IEEE, Grenoble, France (Jul 2010),
  \url{http://ieeexplore.ieee.org/document/5558634/}

\bibitem{rakamaricSMACKDecouplingSource2014}
Rakamari{\'c}, Z., Emmi, M.: {{SMACK}}: {{Decoupling}} source language details
  from verifier implementations. In: International {{Conference}} on {{Computer
  Aided Verification}}. pp. 106--113. Springer (2014)

\bibitem{wolfYosysaFreeVerilog2013}
Wolf, C., Glaser, J., Kepler, J.: Yosys-a free {{Verilog}} synthesis suite. In:
  Proceedings of the 21st {{Austrian Workshop}} on {{Microelectronics}}
  ({{Austrochip}}). p.~97 (2013)

\bibitem{yuReasoningMLIRSemantics2023}
Yu, P.: Reasoning about {{MLIR Semantics}} through {{Effects}} and
  {{Handlers}}. In: Proceedings of the 32nd {{ACM SIGSOFT International
  Symposium}} on {{Software Testing}} and {{Analysis}}. pp. 1552--1554. ACM,
  Seattle WA USA (Jul 2023),
  \url{https://dl.acm.org/doi/10.1145/3597926.3605239}

\end{thebibliography}
\begin{appendix}

\section{Generic MLIR Syntax}
\label{app:generic-mlir-syntax}

\begin{figure}[ht]
\scalebox{0.9}{
\begin{minipage}{\linewidth}
\[\begin{aligned}
&\text{TopLevel}&
\syn{toplevel} \Coloneqq &\ \scmd{(} \syn{operation} \scmd{$\mid$} \syn{type-alias-def} \scmd{$\mid$} \syn{attr-alias-def} \scmd{)*}
\\
&\text{Operation}&
\syn{operation} \Coloneqq &\ \scmd{(} \syn{op-results}\ \tok{=} \scmd{)?}\ \syn{str}\ \tok{(} \scmd{(}\syn{value-uses}\scmd{)?} \tok{)}\ \scmd{(}\tok{[} \syn{successors} \tok{]}\scmd{)?}\
\scmd{(}\tok{<} \syn{dict-prop} \tok{>}\scmd{)?}
\\ 
& &
&\ \scmd{(}\tok{(} \syn{regions} \tok{)}\scmd{)?}\ \scmd{(} \tok{\{} \scmd{(}\syn{dict-attr}\scmd{)?} \tok{\}}\scmd{)?}\ \tok{:}\ \syn{fun-type}\ \scmd{(}  \syn{loc-attr} \scmd{)?}
\\
& &
\syn{op-results} \Coloneqq &\ \syn{op-result}\ \scmd{(} \tok{,}\ \syn{op-result} \scmd{)*} \hspace{20pt} \syn{op-result} \Coloneqq \ \tok{\%}\syn{id}\ \scmd{(} \tok{:}\ \syn{int} \scmd{)?}
\\
& &
\syn{value-uses} \Coloneqq &\ \syn{value-use}\ \scmd{(} \tok{,}\ \syn{value-use} \scmd{)*} \hspace{14pt} \syn{value-use} \Coloneqq \ \tok{\%}\syn{id}\ \scmd{(} \tok{\#}\ \syn{int} \scmd{)?}
\\
& &
\syn{successors} \Coloneqq &\ \syn{successor}\ \scmd{(} \tok{,}\ \syn{successor} \scmd{)*} \hspace{13pt} \syn{successor} \Coloneqq \ \tok{\^}\syn{id}\ \scmd{(}\tok{:}\ \tok{(} \syn{block-args} \tok{)} \scmd{)?}
\\
&\text{Region}& \syn{regions} \Coloneqq &\ \syn{region}\ \scmd{(} \tok{,}\ \syn{region} \scmd{)*} \hspace{51pt} \syn{region} \Coloneqq \ \tok{\{} \syn{operation}^\scmd{+}\ \syn{block}\scmd{*} \tok{\}}
\\
&\text{Block}& \syn{block} \Coloneqq &\ \syn{block-label}\ \syn{operation}^\scmd{+} \hspace{16pt} \syn{block-label} \Coloneqq \ \tok{\^}\syn{id} \ \scmd{(} \tok{(} \syn{block-args} \tok{)} \scmd{)?}\ \tok{:}
\\
& &
\syn{block-args} \Coloneqq &\ \scmd{(} \tok{\%}\syn{id}\ \tok{:}\ \syn{type} \scmd{(} \tok{,}\ \tok{\%}\syn{id}\ \tok{:}\ \syn{type} \scmd{)*} \scmd{)?}
\\
&\text{Type}&
\syn{type-alias-def} \Coloneqq &\ \syn{type-alias} \ \tok{=} \ \syn{type} \hspace{38pt} \syn{type-alias} \Coloneqq \ \tok{!}\syn{bare-id}
\\
& & \syn{func-type} \Coloneqq  &\ \scmd{(} \syn{type} \scmd{$\mid$} \tok{(} \syn{types} \tok{)} \scmd{)} \tok{->} \scmd{(} \syn{type} \scmd{$\mid$} \tok{(} \syn{types} \tok{)} \scmd{)} \hspace{4pt} \syn{types} \Coloneqq \ \syn{type}\ \scmd{(} \tok{,}\ \syn{type} \scmd{)*}
\\
& &
\syn{type} \Coloneqq &\ \syn{type-alias} \scmd{$\mid$} \syn{dialect-type} \scmd{$\mid$} \syn{builtin-type}
\\
&\text{Attribute}&
\syn{attr-alias-def} \Coloneqq &\ \syn{attr-alias}\ \tok{=}\ \syn{attr-value} \hspace{18pt} \syn{attr-alias} \Coloneqq \ \tok{\#}\syn{bare-id}
\\
& &
\syn{dict-attr} \Coloneqq &\ \syn{attr-entry} \ \scmd{(} \tok{,}\ \syn{attr-entry} \scmd{)*} \hspace{6pt} \syn{attr-entry} \Coloneqq \ \scmd{(} \syn{id} \scmd{$\mid$} \syn{str}  \scmd{)} \ \tok{=} \ \syn{attr-value} 
\\
& & \syn{loc-attr} \Coloneqq &\ \tok{loc} \tok{(} \syn{location} \tok{)}
\\
& &
\syn{attr-value} \Coloneqq &\ \syn{attr-alias} \scmd{$\mid$} \syn{dialect-attr} \scmd{$\mid$} \syn{builtin-attr} \hspace{7.5em} 
\\
\end{aligned}\]
\end{minipage}
}
\caption{Syntax for generic MLIR. Note that this figure omits some syntax definitions (e.g., \syn{builtin-attr} --- the builtin attributes of MLIR) for clarity. Indeed, our formalization consists of the complete syntax.}
\end{figure}

\section{List of Formalized Operations}
\label{app:list-of-operations}

\begin{table}[h]
\scriptsize
\centering
\renewcommand{\arraystretch}{1.5}
\setlength\tabcolsep{0.5em}
\caption{Operation list of \textsf{comb} dialect}
\begin{tabular}{r >{\ttfamily}l m{8cm}}
\hline
No. & \textnormal{Operation Name} & Description \\
\hline
1 & comb.add & Integer addition for a variadic list of operands together \\
2 & comb.and & Logical conjunction for a variadic list of operands together \\
3 & comb.concat & Concatenate a variadic list of operands together. \\
4 & comb.divs & Signed integer division \\
5 & comb.divu & Unsigned interger division \\
6 & comb.extract & Extract a range of bits into a smaller value, lowBit specifies the lowest bit included. \\
7 & comb.icmp & Compare two integer values \\
8 & comb.mods & Signed interger modulo \\
9 & comb.modu & Unsigned interger modulo \\
10 & comb.mul & Integer multiplication for a variadic list of operands together \\
11 & comb.mux & Return one or the other operand depending on a selector bit \\
12 & comb.or & Logical disjunction for a variadic list of operands together \\
13 & comb.parity & Logical parity operation for an integer \\
14 & comb.replicate & Concatenate the operand a constant number of times \\
15 & comb.shl & Logical left shifting \\
16 & comb.shrs & Logical right shifting for signed intergers \\
17 & comb.shru & Logical right shifting for unsigned intergers \\
18 & comb.sub & Integer subtraction \\
19 & comb.truth\_table & Return a true/false based on a lookup table \\
20 & comb.xor & Logical XOR operation for a variadic list of operands together \\
\hline
\end{tabular}
\end{table}

\begin{table}[h]
\scriptsize
\centering
\renewcommand{\arraystretch}{1.2}
\setlength\tabcolsep{0.5em}
\caption{Operation list of \textsf{hw} dialect}
\begin{tabular}{r >{\ttfamily}l m{8cm}}
\hline
No. & \textnormal{Operation Name} & Description \\
\hline
1 & hw.module & Represents a Verilog module \\
2 & hw.hierpath & Represents a path through the hierarchy \\
3 & hw.instance & Create an instance of a module \\
4 & hw.output & HW termination operation \\
5 & hw.bitcast & Reinterpret one value to another value of the same size and potentially different type \\
6 & hw.constant & Produce a constant value \\
7 & hw.enum.cmp & Compare two values of an enumeration \\
8 & hw.enum.constant & Produce a constant enumeration value \\
9 & hw.wire & Assign a name or symbol to an SSA edge \\
10 & hw.aggregate\_constant & Produce a constant aggregate value \\
11 & hw.array\_concat & Concatenate some arrays \\
12 & hw.array\_create & Create an array from values \\
13 & hw.array\_get & Get the value in an array at the specified index \\
14 & hw.array\_slice & Get a range of values from an array \\
15 & hw.struct\_create & Create a struct from constituent parts \\
16 & hw.struct\_explode & Expand a struct into its constituent parts \\
17 & hw.struct\_extract & Extract a named field from a struct \\
18 & hw.struct\_inject & Inject a value into a named field of a struct \\
19 & hw.union\_create & Create a union with the specified value \\
20 & hw.union\_extract & Get the value of a union, interpreting it as the type of the specified member field \\
\hline
\end{tabular}
\end{table}

\begin{table}[h]
\scriptsize
\centering
\renewcommand{\arraystretch}{1.5}
\setlength\tabcolsep{0.5em}
\caption{Operation list of \textsf{seq} dialect}
\begin{tabular}{r >{\ttfamily}l l}
\hline
No. & \textnormal{Operation Name} & Description \\
\hline
1 & comb.add & Integer addition for a variadic list of operands together \\
2 & seq.firmem.read\_port & A memory read port \\
3 & seq.firmem.read\_write\_port & A memory read-write port \\
4 & seq.firmem.write\_port & A memory write port \\
5 & seq.firreg & Register with preset and sync or async reset \\
\hline
\end{tabular}
\end{table}

\begin{table}[!h]
\scriptsize
\centering
\renewcommand{\arraystretch}{1.25}
\setlength\tabcolsep{0.5em}
\caption{Operation list of \textsf{sv} dialect}
\begin{tabular}{r >{\ttfamily}l m{8cm}}
\hline
No. & \textnormal{Operation Name} & Description \\
\hline
1 & sv.always & \verb|always @| block in SystemVerilog \\
2 & sv.alwayscomb & \verb|alwayscomb| block in SystemVerilog \\
3 & sv.alwaysff & \verb|alwaysff @| block with optional reset \\
4 & sv.array\_index\_inout & Index an inout memory to produce an inout element \\
5 & sv.assert & Immediate assertion statement \\
6 & sv.assert.concurrent & Concurrent assertion statement \\
7 & sv.assign & Continuous assignment \\
8 & sv.assume & Immediate assume statement \\
9 & sv.assume.concurrent & Concurrent assume statement \\
10 & sv.bind & Indirect instantiation statement \\
11 & sv.bpassign & Blocking procedural assignment \\
12 & sv.case & \verb|case (cond)| block in SystemVerilog \\
13 & sv.constantX & A constant of value \verb|x| \\
14 & sv.constantZ & A constant of value \verb|z| \\
15 & sv.cover & Immediate cover statement \\
16 & sv.cover.concurrent & Concurrent cover statement \\
17 & sv.error & "error" severity message task \\
18 & sv.exit & "exit" system task \\
19 & sv.fatal & "fatal" severity message task \\
20 & sv.finish & "finish" system task \\
21 & sv.for & for loop in SystemVerilog \\
22 & sv.force & Force procedural statement in SystemVerilog \\
23 & sv.fwrite & fwrite statement \\
24 & sv.generate & A \verb|generate| block \\
25 & sv.generate.case & A \verb|case| statement inside of a generate block \\
26 & sv.if & \verb|if (cond)| block \\
27 & sv.ifdef & \verb|ifdef MACRO| block \\
28 & sv.ifdef.procedural & \verb|ifdef MACRO| block for procedural regions \\
29 & sv.indexed\_part\_select & Read several contiguous bits of an int type \\
30 & sv.indexed\_part\_select\_inout & Address several contiguous bits of an inout type \\
31 & sv.info & "info" severity message task \\
32 & sv.initial & \verb|initial| block \\
33 & sv.logic & Define a logic \\
34 & sv.macro.decl & SystemVerilog macro declaration \\
35 & sv.macro.def & SystemVerilog macro definition \\
36 & sv.macro.ref & Expression to refer to a SystemVerilog macro \\
37 & sv.macro.ref.se & Expression to refer to a SystemVerilog macro \\
38 & sv.ordered & A region which guarantees to output statements in order \\
39 & sv.passign & Nonblocking procedural assignment \\
40 & sv.read\_inout & Get the value of from something of inout type as the value itself \\
41 & sv.readmem & Load a memory from a file in either binary or hex format \\
42 & sv.reg & Define a new register in SystemVerilog \\
43 & sv.release & Release procedural statement \\
44 & sv.stop & "stop" system task \\
45 & sv.struct\_field\_inout & Create an subfield inout memory to produce an inout element. \\
46 & sv.warning & "warning" severity message task \\
47 & sv.wire & Define a new wire \\
48 & sv.xmr & Encode a reference to a non-local net. \\
49 & sv.xmr.ref & Encode a reference to something with a hw.hierpath. \\
50 & sv.verbatim & Verbatim opaque text emitted inline. \\
51 & sv.verbatim.expr & Expression that expands to a value given SystemVerilog text \\
52 & sv.verbatim.expr.se & Expression that expands to a value given SystemVerilog text \\
\hline
\end{tabular}
\end{table}

\FloatBarrier
\section{Comparison between Kimulator and Verilator}
\label{app:comparison-kimulator-varilator}

\begin{figure}[h]
\vspace{-7pt}
    \centering
    \includegraphics[width=0.8\linewidth]{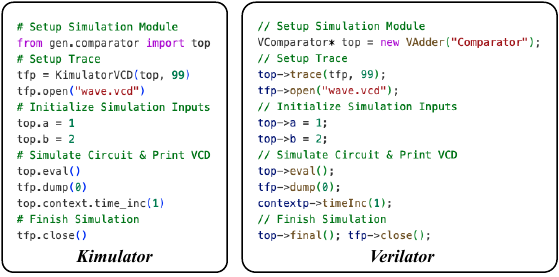}
    \vspace{-3pt}
    \caption{Comparative simulation illustration for Kimulator and Verilator.}
    \label{fig:kimulator-example}
    \vspace{-18pt}
\end{figure}

\end{appendix}

\end{document}